\def\aa{{A\&A}}
\def\aas{{A\&AS}}

\def\an{{AN}}
\def\annrev{{ARA\&A}}
\def\apj{{ApJ}}
\def\apjs{{ApJS}}

\def\mnras{{MNRAS}}
\def\nat{{Nature}}

\def\asec{$^{\prime\prime}$}
\def\amin{$^\prime$}

\documentclass[cup5b]{caps}
\usepackage{graphicx}
\usepackage{amssymb}
\usepackage{latexsym}
\usepackage{ociwsymp3}   

\HeadText{A. C. Edge}

\begin{document}

\pagenumbering{arabic}

\author[]{ALASTAIR C. EDGE\\ Institute for Computational Cosmology, 
University of Durham}

\chapter{X-ray Surveys of Low-redshift \\ Clusters}

\begin{abstract}

The selection of clusters of galaxies through their X-ray emission
has proved to be an extremely powerful technique over the past four
decades. The growth of X-ray astronomy has provided the community
with a steadily more detailed view of the intracluster medium in
clusters. In this review I will assess how far X-ray surveys of
clusters have progressed and how far they still have to travel.

\end{abstract}

\section{Introduction}

The principal baryonic component of clusters of galaxies is
diffuse gas held in hydrostatic equilibrium in the 
gravitational potential of the cluster. This gas is
hot (10$^{7}-10^{8}$ K), relatively dense 
(10$^{-4}-10^{-2}$ atom cm$^{-3}$), 
and enriched with heavy elements 
(e.g., Fe of 0.3 Solar abundance). This combination results
in significant X-ray emission through thermal bremsstrahlung
radiation. Detailed X-ray observations of clusters can provide
us with accurate total mass measurements, clues to the
merger history of clusters, and a chemical record of the
supernova ejecta that polluted the intracluster medium
during the formation of the stars in the member galaxies.

The X-ray emission from clusters can also be exploited to
select clusters irrespective of their member galaxies. While
the optical selection of clusters is well established and
understood, there are potential problems with projection
and the imperfect scaling of the galaxy population to
total cluster mass that make independent selection methods
attractive. 

There are four key considerations for any X-ray survey:

\begin{itemize}
\item{{\bf Spatial resolution} To capitalize on the extended nature of 
the X-ray emission in clusters, it is important to
have sufficient spatial resolution to differentiate
clusters from most other pointlike X-ray sources
(i.e., stars and AGNs). On the other hand, the most
nearby, diffuse clusters can be missed in the same
way low-surface brightness galaxies may be missed in 
optical surveys. }

\item{{\bf Spectral resolution} Each class of X-ray
source has a distinctive spectral signature (e.g., black
body for white dwarfs), so information on the X-ray
spectrum of each source can aid identification. The
thermal nature of cluster spectra (temperatures mostly
greater than 2~keV) give clusters relatively flat
soft X-ray spectra (making them distinctive in the {\it ROSAT} 
survey), but the overall spectral shape is similar to most
unabsorbed AGNs. Therefore, definitive cluster identification
from spectral data alone requires many photons ($>$1,000),
which is only feasible for the brightest detections 
(see Nevalainen et al. 2001 for an example using {\it XMM-Newton}).}

\item{{\bf Flux limit of the survey} This is a particularly
important factor for cluster surveys as the flux-limited
nature of X-ray samples translates to a selection
over a wide range in redshifts, since the most luminous
objects are being selected from a very much larger volume than
the least luminous ones. This has its advantages but
requires careful analysis. This is illustrated in Figure~1.1,
where the X-ray luminosity is plotted against redshift for
a variety of samples described later in the text.}

\item{{\bf Area of sky surveyed}  In the ideal survey at any
wavelength the aim is all-sky coverage. This has been 
achieved several times in X-ray astronomy with the earliest
X-ray satellites scanning with collimated proportional counters
($\sim 1^\circ$ resolution) and {\it ROSAT} with a soft X-ray
imaging telescope ($\sim 1^\prime$ resolution). This wide coverage
comes at the expense of depth [the {\it ROSAT} All-Sky Survey 
reaches $\sim 10^{-12}$ erg~s$^{-1}$cm$^{-2}$ (0.5--2~keV)].
The alternative survey strategy is to select serendipitous
sources in pointed imaging observations, as pioneered by the
Extended {\it Einstein} Medium Sensitivity Survey (EMSS; Gioia et al. 1990).
This allows much deeper surveys but at the expense of
the area (and hence total volume) covered.}
\end{itemize}

\begin{figure*}[t]
\includegraphics[width=1.00\columnwidth,angle=0,clip]{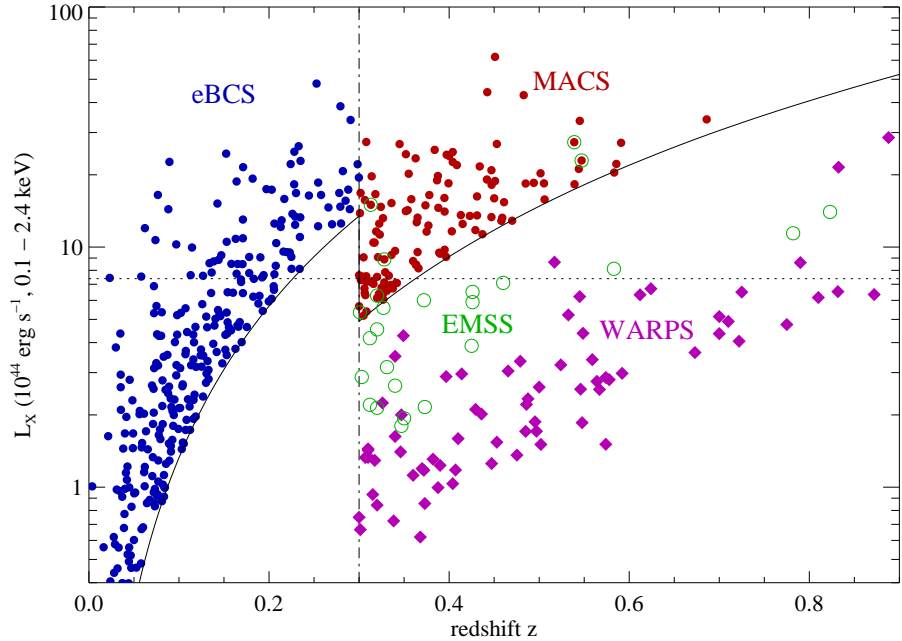}
\vskip 0pt \caption{
The X-ray luminosity (0.1--2.4~keV) plotted against
redshift for four X-ray samples: EMSS (Gioia et al. 1990), a 
serendipitous sample from {\it Einstein}; eBCS (Ebeling et al. 2000), a
shallow, wide RASS survey; MACS (Ebeling et al. 2001), a deeper, wide RASS 
survey, and WARPS (Perlman et al. 2002) a much deeper serendipitous 
{\it ROSAT} survey. 
\label{Lx-redshift}}
\end{figure*}

There are a number of cluster properties that can be
used to constrain the nature and evolution of clusters.

\begin{itemize}

\item{{\bf Temperature} - For gas in hydrostatic equilibrium
(which appears to hold for the majority of the volume of a cluster) 
the gas temperature and density can be used to directly determine the
cluster mass.}

\item{{\bf Elemental abundances} The X-ray spectrum of clusters contains
lines from a number of heavy elements. Most prominent of these is the
iron 6.7~keV line. The inferred abundance ratios from X-ray spectra
of O, S, Si and Fe can be used to determine the dominant supernova
type (Loewenstein \& Mushotzky 1996). }

\item{{\bf Surface brightness profiles} The distribution of gas in
a cluster has a very significant effect on the total X-ray luminosity
of the cluster, given that the intensity of emission is proportional to
the density squared. Clusters with compact, dense cores (i.e., cooling flows;
see Fabian 1994) are much more luminous that more extended clusters of
the same measured X-ray temperature (Fabian et al. 1994) and can have
an effect on the detection probability in X-ray surveys (Pesce et al. 1990).
Also, the
recent discovery of strong density discontinuities, termed ``cold fronts''
(Markevitch et al. 2000; Mazzotta et al. 2001) 
has highlighted the impact of past mergers on the intracluster medium. These 
%XXX no ICM
factors make obtaining high-quality, high-resolution X-ray imaging
a vital element of cluster studies.}

\end{itemize}

Each of these requires either dedicated pointed observations 
or a survey drawn from the brightest serendipitous detections in pointed
observations. The former is a relatively slow process requiring
time allocation committees to put substantial resources into
programs to observe ``complete'' samples. The latter is
very slow given the area covered by sufficiently deep X-ray 
observations.

For the purposes of this review I will define ``low redshift'' as
$z<0.5$ and treat any paper presenting any new X-ray detection of
cluster as a ``survey.''

In my talk I used the yardstick of exponential growth to 
judge progress in known numbers of X-ray emitting clusters
which I modestly named ''Edge's Law.'' This holds that
for every decade of X-ray astronomy the number of clusters
detected increases by an order of magnitude. I would like
to stress that this was a narrative device and not a
serious bid for future surveys in itself. That said, the
rapid progress in cluster research in the past decades
does require us to stand back and assess it as part of
a larger picture.

\section{An Historical Perspective}

I would like to continue this review in a similarly
light-hearted vein while giving the reader as comprehensive
review as possible of X-ray cluster surveys. 

\subsection{In the Beginning...}

X-ray astronomy began on the 18th of June 1962 with
the detection of the X-ray background and Sco-X1 by
a sounding rocket experiment (Giacconi et al. 1962).
For shorthand in this review, this will be denoted as
0 Anno Giacconi (AG), and subsequent events will be
quoted in these units.

During the first three years of sounding rocket
experiments from 0~AG several cluster detections
were in dispute
[e.g., Coma claimed by Boldt et al. (1966) and 
discounted by Friedman \& Byram (1967)], 
so the first unambiguous cluster detection came in
4~AG when Byram, Chubb, \& Freidman (1966) detected M87/Virgo.

The numerous sounding rocket campaigns that occurred
between 1962 and 1975 resulted in several more
cluster detections (e.g., Perseus; Fritz et al. 1971) and the discovery that
the X-ray emission in Coma was extended (Meekins et al. 1971).
Unfortunately the collecting area and exposure time
of these experiments ruled out the detection of 
all but the brightest few clusters. 

On 12th December 1970 the first X-ray satellite, {\it UHURU},
was launched. The two-year lifetime of the mission
allowed the whole sky to be scanned many times,  
producing the first true X-ray survey. The fourth and final
{\rm UHURU} catalog (4U; Forman et al. 1978) contains 52 clusters, several of
which that were not known to be clusters at the time.

\subsection{The End of the First Age of X-ray Astronomy?}

{\it UHURU} was the first of a number of increasingly more
complicated experiments that allowed
further surveys and dedicated pointed observations.
Most notable of these was {\it Ariel-V} (Cooke et al. 1978), 
which made the first iron line detection in a
cluster (Mitchell et al. 1976).

The final mission in this series, {\it HEAO-1}, made the
deepest X-ray survey (Piccinotti et al. 1982),  which was the mainstay of
X-ray astronomy for the two decades that followed.
So at the end of this exciting period of X-ray astronomy,
how well does ``Edge's Law'' stand up? At 17~AG a total
of 95 clusters are known, which is well above the 
50 required by this point.

\section{X-ray Imaging Begins with {\it Einstein} }

The launch of the first imaging X-ray satellite, {\it Einstein}, had
a profound impact on our understanding of clusters.

\subsection{Detailed Imaging and Spectra}

The Imaging Proportional Counter (IPC) and High Resolution Imager
(HRI) provided images of unprecedented quality (up to 5\asec\ FWHM).
These two instruments provided a great deal of detailed
information on individual clusters from targeted observations
(Fabian et al. 1981; Jones \& Forman 1984, 1999; Stewart et al. 1984; 
White, Jones, \& Forman 1997). 
Most of the observations were of Abell clusters or other
optically selected clusters, but radio galaxies in clusters were
also targets. Given the nature of the targeted observations it is
not possible to derive any stringent limits on the statistical
properties of clusters, but a luminosity function was derived
for Abell clusters (Burg et al. 1994). 

{\it Einstein} also carried the first semi-conductor detector (the
forerunner to today's CCDs), the Solid State Spectrometer (SSS) and
a deployable Bragg Crystal, the Focal Plane Crystal Spectrometer (FPCS).
Both of these instruments provided important results for clusters
(Canizares et al. 1979; White et al. 1991), but were
not used systematically.

\subsection{The EMSS}

The {\it Einstein} survey that has had the most impact on cluster research is 
undoubtedly the Extended {\it Einstein} Medium Sensitivity Survey (EMSS).
By combining most of the serendipitous detections from IPC observations, it 
was possible to survey 980~$\Box^{\circ}$ and detect 99 clusters (Stocke et 
al. 1991). This sample has since been the basis for a
great deal of work at all wavelengths (Donahue, Stocke, \& Gioia 1992; 
Le F\'evre et al. 1994; Carlberg et al. 1997; Luppino et al. 1999)

\section{The X-ray Dark Ages}

The 1980's were a period of relative calm in X-ray astronomy.
The only satellites launched between 1980 and 1990 were the ESA mission 
{\it EXOSAT} and two Japanese missions, {\it Tenma} and {\it Ginga}.
All three of these satellites were designed for targeted
follow-up of known objects, and only {\it EXOSAT} had any 
imaging capabilities. {\it EXOSAT} and {\it Ginga} contributed a
significant number of accurate cluster temperature and
iron abundance measurements (the data from {\it EXOSAT} making
my thesis) but very few ``new'' detections.

This lull in proceedings did allow the previous scanning
and {\it Einstein} surveys to be collated, and a complete
sample of the brightest 55 clusters was compiled
(Lahav et al. 1989). This sample has been used
to determine the first cluster temperature function
(Edge et al. 1990), the first correlation function
from an X-ray sample (Lahav et al. 1989), and the 
fraction of cooling flows (Edge, Stewart, \& Fabian 1992;
Peres et al. 1998).

This ``free-wheeling'' in X-ray surveys leaves the number
of clusters in 28~AG at 300, well short of the 630 required
for my exponential growth. This gap did not last for long....

The German/UK/US satellite {\it ROSAT} was launched on the 1st of June
1990 (27.95~AG). The wide-field, soft X-ray imaging telescope
of {\it ROSAT} was used to conduct a 6-month scanning survey of the whole sky,
from August 1990 to February 1991, which detected in excess of 100,000
sources to a flux limit of $(0.3-1)\times 10^{-12}$ erg~s$^{-1}$cm$^{-2}$ (0.1--2.4~keV)
(depth depending on position). From March 1991 to December 1998, 
{\it ROSAT} performed a series of pointed observations with the
PSPC and HRI detectors. These observations targeted many known
and recently detected clusters, as well as detecting a great many
clusters serendipitously.

\begin{table*}[t]
\centering
\caption{{\it ROSAT} Survey Samples}
\begin{tabular}{llccc}
\hline
Survey  & Identification & Flux Limit & Area & Number  \\
        & Paper     & (erg~s$^{-1}$cm$^{-2}$) & ($\Box^\circ$) & Published?  \\
 & & & &  \\ \hline
XBACS   & Abell clusters          & 5.0$\times 10^{-12}$ & All-sky &  276   \\
        & Ebeling et al. (1996)   & (0.1--2.4 keV) & & Y \\
BCS     & Abell, Zwicky, extended & 4.5$\times 10^{-12}$ & 13,578       &  199   \\
        & Ebeling et al. (1998)   & (0.1--2.4 keV) & & Y \\
RASS1BS & Abell, extended         & 3-4$\times 10^{-12}$   & 8,235       &  130   \\
        & de Grandi et al. (1999) & (0.5--2.0 keV) & & Y \\
Ledlow  & Abell $z<0.09$          & none                   & 14,155 & 294 \\
        & Ledlow et al. (1999)    &                        & & N \\
eBCS    & Abell, Zwicky, extended & 3.0$\times 10^{-12}$ & 13,578     &  299   \\
        & Ebeling et al. (2000)   & (0.1--2.4 keV) & & Y \\
HiFLUGS & All                     & 20$\times 10^{-12}$    & 27,156 &   63   \\
        & Reiprich \& B\"ohringer (2002)   & (0.1--2.4 keV) & & Y \\
NORAS   &        extended         & 3.0$\times 10^{-12}$ &  13,578    &  378   \\
        & B\"ohringer et al. (2000)   & (0.1--2.4 keV) & & Y \\
NEP     & multiple                & 0.03$\times 10^{-12}$ & 80.7      &  64     \\
        & Gioia et al. (2001)   & (0.5--2.0 keV) & & Y \\
CIZA    & CCD imaging, $|b|<20^\circ$ & 5$\times 10^{-12}$ & 14,058 & 73   \\
        & Ebeling, Mullis, \& Tully (2002)   & (0.1--2.4 keV) & & Y \\
SGP     & optical plates scans    & 3.0$\times 10^{-12}$   & 3,322       &  112   \\
        & Cruddace et al. (2002)  & (0.1--2.4 keV) & & Y \\
MACS    & multiple,  $z>0.3$      & 1.0$\times 10^{-12}$ & 22,735  &  120   \\
%XXX moved  $z>0.3$
        & Ebeling et al. (2001)   & (0.1--2.4 keV) & & N \\
REFLEX  & multiple                & 3.0$\times 10^{-12}$ & 13,905       &  452    \\
        & B\"ohringer et al. (2001)   & (0.1--2.4 keV) & & N \\ \hline
\end{tabular}
\end{table*}

\subsection{{\it ROSAT} All-Sky Survey}

The {\it ROSAT} All-Sky Survey (RASS) is a resource that has still yet to 
fully exploited 12 years after it was completed. A number of
coordinated cluster surveys were embarked
upon as soon as the RASS ended. The understandably tight control over the RASS
data release and the time-consuming nature of the optical follow-up
of the clusters has meant a significant lag in the publication of 
these samples. Table 1.1 lists a representative set of
RASS cluster surveys, both published and unpublished.
There are other RASS studies containing clusters (e.g., 
the RBS, a complete sample of all RASS sources to a count rate limit of 0.2 
PSPC count~s$^{-1}$; Schwope et al. 2000),
and studies of groups (e.g., RASSCALS;  Mahdavi et al. 2000) and 
Hickson compact groups (Ebeling, Voges, \& B\"ohringer 1994). Table 1.1 will
be added to in the next few years by NORAS-2, REFLEX-2, and eMACS, 
which will extend each of the existing surveys to lower fluxes, 
but these will be reaching close to the intrinsic sensitivity limit of
the majority of the RASS.

The wide variety of selection criteria, detection methods and
areas covered are clear from Table 1.1. However, each of the 
larger samples (BCS, 1BS, Ledlow, REFLEX, and NEP) agree in their
derived X-ray luminosity functions (Ebeling et al. 1997; de~Grandi et al. 
1999; Ledlow et al. 1999; Gioia et al. 2001; B\"ohringer et al. 2002), 
so these differences do not greatly affect the samples. 

One important difference in the principal RASS samples
is that one set (XBACS, BCS, and eBCS) is based on a selection
using a Voronoi-Percolation-and-Tesselation (VTP) technique (Ebeling \& 
Wiedenmann 1993) and the other (1BS, SGP, NORAS, and REFLEX) is 
based on a growth curve analysis (GCA) technique (B\"ohringer et al. 2001).
The flux results from both methods agree within the errors, 
but only VTP acts as a detection algorithm, as the GCA
method requires a set of input positions of potential clusters.
This difference is relevant only for the most nearby, extended
sources, which are not detected by detection algorithms tuned
to search for point sources. VTP will reliably detect these,
but, through lack of access to the full RASS data set, it was not
run over the full sky during the compilation of the BCS. 
With all RASS data now in the public domain, this is now
possible in principle.

The optical follow-up of clusters at redshifts above 0.3
requires additional optical imaging, as archival photographic
plate material is too shallow to reliably detect cluster
members. At the brighter flux limits (5$\times 10^{-12}$ erg~s$^{-1}$cm$^{-2}$)
there are relatively few of these distant clusters [e.g., two in the BCS
and RXJ1347--11 ($z=0.45$) in RASS1BS], but this number increases
with decreasing flux limit (e.g., there are seven in the eBCS). 
With these higher redshift, X-ray luminous clusters in
mind, Harald Ebeling and I have searched the RASS-BSC sample
(Voges et al. 1999) for $z>0.3$ clusters using the UH~2.2~m telescope
to a flux limit of $10^{-12}$ erg~s$^{-1}$cm$^{-2}$, 
creating the MAssive Cluster Survey (MACS; Ebeling, Edge, \& Henry 2001).
To date, the sample contains 120 clusters with very few candidates
left for imaging. The MACS sample has been extensively
followed up at all wavelengths, including complete $VRI$ imaging 
with the UH~2.2~m, multi-object spectroscopy
with Keck, Gemini and CFHT, deep, wide-area imaging with Subaru/SUPRIMECAM,
VLA imaging (Edge et al. 2003), Sunyaev-Zel'dovich observations
(LaRoque et al. 2003), {\it Chandra} observations, and Cycle 12 {\it HST}/ACS
imaging. This sample represents more than an order of magnitude improvement
in the number of distant, X-ray luminous clusters known (i.e., two 
EMSS clusters with $z>0.4$, $L_x>10^{45}$~erg~s$^{-1}$, compared to
39 in MACS).

\begin{table*}[t]
\centering
\caption{{\it ROSAT} Serendipitous Samples}
\begin{tabular}{llccc}
\hline
Survey  & Selection & Flux Limit & Area & Number  \\
& Paper   & ($10^{-14}$ erg~s$^{-1}$cm$^{-2}$) & ($\Box^\circ$)& Published?  \\
 & & & &  \\ \hline
RIXOS         & CCD imaging & 3.0 & 15.8 & 25 \\
              & Mason et al. (2000) & (0.5--2.0 keV)  & & Y \\
WARPS-I       & CCD imaging & 6.5 & 14.1 & 25 \\
              & Perlman et al. (2002) & (0.5--2.0 keV)  & & Y \\
160sq.deg.    & extent      & 3.0 & 158 & 203 \\
              & Vikhlinin et al. (1998) & (0.5--2.0 keV) & & Y \\
SHARC-S       & extent      & 3.9 & 17.7 & 16 \\
              & Collins et al. (1997) & (0.5--2.0 keV) & & N \\
Bright SHARC  & extent      & 16.3 & 179 &  37 \\
              & Romer et al. (2000)   & (0.5--2.0 keV) & & Y \\
RDCS          & extent      & 3.0 & 50 & 103 \\
              & Borgani et al. (2001) & (0.5--2.0 keV) & & N \\
ROXS          & CCD imaging & 2.0 & 4.8 & 57 \\
              & Donahue et al. (2002)   & (0.5--2.0 keV) & & Y \\
BMW           & extent      & $\sim 10$ & $\sim 300$ & $\sim$100 \\
              & Lazzati et al. (1999) & (0.1--2.4 keV) & & N \\
XDCS          & CCD imaging & 3.0 & 11.0 & 15 \\
              & Gilbank et al. (2003) & (0.5--2.0 keV)  & & N \\
WARPS-II      & CCD imaging & 6.5 & 73 & 150 \\
              & Jones et al., in prep. & (0.5--2.0 keV)  & & N \\ \hline
\end{tabular}
\end{table*}

\subsection{{\it ROSAT} Pointed Observations}

The large field of view of the PSPC detector made it very
efficient at detecting serendipitous X-ray sources within 15$^\prime$
of the pointing position of the telescope. Given the substantial
numbers of relatively deep observations during the {\it ROSAT} Pointed
Phase an area of well over 500$\Box^\circ$ at high Galactic latitude 
has been covered by the central region of the PSPC with more than 10~ks 
exposure.  This area is significantly reduced, 
as some targets are not suitable for serendipitous searches (e.g.,
nearby clusters, nearby galaxies, globular clusters, etc.), but
the majority has been used in a series of surveys (listed in Table 1.2).
As with the RASS samples, the selection strategies differ between
surveys, but the results from each survey agree. For instance,
the requirement of significant source extent used by SHARC certainly
eases source selection, but at the potential loss of the most
distant and/or compact, cooling flow clusters. These effects
do not appear to have any great impact on the results.

The majority of the clusters selected in these surveys are
relatively nearby ($z=0.15-0.3$) and of low X-ray
luminosity ($L_x=10^{43-44}$ erg~s$^{-1}$; 0.5--2~keV), but
a few distant, luminous clusters are found (Ebeling et al. 2000, 2001), and
in the deepest of the {\it ROSAT} serendipitous sample, RDCS (Rosati et al. 
1998), there are several candidate clusters at $z>1$ (Borgani et al. 2001). 

The PSPC instrument eventually ran out of gas in mid-1994, but {\it ROSAT} 
continued to make observations with the HRI. While this
instrument was less sensitive than the PSPC and  covered a 
smaller area of sky, the excellent spatial resolution
it provided has been used very effectively by the Brera Observatory group
in the BMW survey (Lazzati et al. 1999; Panzera et al. 2003). While the 
combination of sensitivity and total area covered by the BMW survey will never 
match that of PSPC surveys (e.g., the 160 square degree survey; Vikhlinin 
et al. 1998), it does provide an important reliability test.

The full potential of the {\it ROSAT} pointed phase has yet to be tapped
as all existing surveys have been restricted to the central 15\amin--20\amin\ 
radius where the point-spread function is best. While the flux sensitivity is 
poor in the outer parts of the detector, the brighter 
($f_x>10^{-13}$ erg~s$^{-1}$~cm$^{-2}$) sources can easily be detected.
As noted above, the most time consuming part of the follow-up of X-ray
selected cluster candidates at $z>0.3$ is the deeper optical imaging
required. This is exacerbated at lower X-ray fluxes by the fainter
optical counterparts of all X-ray counterparts. The combination of
the low-resolution X-ray imaging with deeper multicolor panoramic
surveys such as SDSS, UKIDSS, and RCS2 will provide a ``free'' resource
to identify cluster counterparts to these {\it ROSAT} sources. 

So, how successful has {\it ROSAT} been overall in harvesting clusters?
The RASS samples published or about to be published account for a
total of around 1,200 new clusters, with a further 500--1,000 at lower fluxes.
Add these to the serendipitous detections, $\sim$500 in the central
region of the PSPC (of which $\sim$250 are published), $>$1,500 in the outer
regions, and $\sim$300 clusters in the HRI. Therefore, the total
after the {\it ROSAT} mission is $\sim$4,000 (when in 36.5~AG 4,500 would
be required). It is worth noting that the majority of these clusters have
yet to be identified and many may never be.

\section{The Middle Age of X-ray Astronomy?}

The arrival of the third decade of X-ray astronomy midway
through the {\it ROSAT} mission coincided with advances in CCD
detector technology that have allowed a vast increase in the
power of X-ray spectroscopy. These advances, coupled with
nested-mirror systems, mark a clear maturing in the field
and a move away from large samples of objects with limited
information to limited samples with very detailed information.
This is an inevitable progression that emerging disciplines
experience, radio astronomy being a prime example. With the
advance of aperture synthesis, radio astronomers 
in $\sim$35 AJ (Anno Jansky) could obtain insights
into the nature of individual sources and the focus
moved away from surveys. In the past decade, radio astronomy
has turned back to surveys (NVSS, FIRST, WENSS, 4MASS), and this
will happen in X-ray astronomy (but hopefully in less than
25 years time!).

This trend for more detailed study has had a huge impact on
cluster research, and the need for spatially resolved 
spectroscopy of clusters was apparent from the first X-ray
detections of clusters. The nature of cluster
surveys has also changed with a greater emphasis on
understanding complete samples in many different
wavelength regimes (e.g., Crawford et al. 1999; Giovannini, Tordi, 
\& Ferretti 1999; Pimbblet et al. 2002). A sample of 200
clusters is a great resource but of little use
without some information about the X-ray temperature, iron
abundance, X-ray surface brightness profile, optical
photometry and spectroscopy, or radio imaging. 
The availability of new optical, near-infrared, and radio surveys
(e.g., SDSS, UKIDSS, NVSS, FIRST) will make the
multiwavelength aspects of these studies much
easier, but the need for further X-ray observations
is hard to avoid.

\subsection{{\it ASCA} Observations}

The first step in this progression was the Japanese-US
satellite {\it ASCA}. The nested, foil-replicated mirrors of
{\it ASCA} resulted in a relatively asymmetric, broad point-spread 
function (2\amin\ FWHM), but the excellent performance
of the SIS CCD detectors provided some very high-quality
spectra for clusters (Mushotzky \& Scharf 1997; Markevitch 1998; 
Fukazawa et al. 2000; Ikebe et al. 2002). 

Over the course of the seven-year pointed phase, {\it ASCA} provided accurate
temperatures and iron abundances for most of the 350 clusters observed. While few complete
samples were observed, the {\it ASCA} data are an excellent
complement to archival {\it ROSAT} observations.
The notable exceptions to this are the flux-limited sample of 
61 {\it ROSAT}-selected clusters (Ikebe et al. 2002) and 
the complete sample of $0.3<z<0.4$ EMSS clusters (Henry 1997) 
from which limits of the evolution of the cluster temperature function 
can be derived.

\subsection{The Unfulfilled Potential of {\it ABRIXAS}}

One of the most disappointing events in X-ray astronomy was
the unfortunate failure of the German satellite {\it ABRIXAS} in
June 1998. Its simple design
and the track record of the team behind {\it ROSAT} 
meant the planned 3-year, all-sky survey {\it ABRIXAS} 
would have had a huge impact on X-ray astronomy. The survey depth envisioned 
of 1.5$\times 10^{-13}$ erg~s$^{-1}$cm$^{-2}$ (0.5--2.0~keV)
and 9$\times 10^{-13}$ (2--12~keV) would have detected
in excess of 20,000 clusters (i.e., more than the
number required to keep pace with exponential growth).

\subsection{{\it Chandra} and {\it XMM-Newton}}

The launch of {\it Chandra} and {\it XMM-Newton} in 1999 has seen
X-ray astronomy reach full maturity. The sub-arcsecond
imaging of {\it Chandra} and unprecedented throughput of
{\it XMM-Newton} have had a profound impact of our
understanding of clusters (e.g., McNamara et al. 2000; Peterson et al. 2001;
Allen, Schmidt, \& Fabian 2002).
The potential for surveys with both satellites is 
largely through serendipitous detections, but several
important pointed surveys are being undertaken.

The only large {\it Chandra} serendipitous survey is
CHamP (Wilkes et al. 2001), which will cover 14$\Box^\circ$ in 5 years and
identify 8,000 X-ray sources of all types, of
which 150--250 will be clusters (which will
all be spatially resolved). The relatively
small number of clusters makes this sample unlikely
to set any strong cosmological constraints, but
it will act as an excellent control sample for
past and future samples to test how spatial
resolution affects detection statistics.

{\it XMM-Newton} has a program similar to CHamP, the XID program that has 
three tiers: faint ($10^{-15}$ erg~s$^{-1}$cm$^{-2}$, 0.5$\Box^\circ$), medium 
($10^{-14}$ erg~s$^{-1}$cm$^{-2}$, 3$\Box^\circ$), and bright 
($10^{-13}$ erg~s$^{-1}$cm$^{-2}$, 100$\Box^\circ$). Again, like CHamP, the 
number of clusters detected in the XID program will be small ($<50$), so from
a purely cluster view-point is not particularly relevant. 
There are currently two dedicated serendipitous cluster surveys.
One expands on the XID programme (Schwope et al. 2003) and 
the other (the X-ray Cluster Survey, XCS, Romer et al. 2001) aims to 
extract all potential cluster candidates
from the {\it XMM-Newton} archive and compile a sample of $>$5,000
clusters from up to 1,000$\Box^\circ$ over the full lifetime of
the satellite. The contrast of XCS to
CHamP and XID illustrates the huge increase in efficiency
when one class of objects is chosen over the study of ``complete'' X-ray
samples or contiguous area X-ray surveys, such as the XMM-LSS (Pierre et al. 2003),
where the number of detected cluster is relatively small.

The principal pointed cluster surveys with {\it Chandra} and/or {\it XMM-Newton} 
target a sample of MACS clusters (Ebeling et al. 2001) with
{\it Chandra} using GTO and GO time (PIs Van Speybroeck and Ebeling),
a sample of REFLEX clusters with {\it XMM-Newton} in GO time (PI B\"ohringer)
and a sample of SHARC clusters with {\it XMM-Newton} in GTO time (PI Lumb).
Each of these projects is designed to determine the cluster
temperature function, but will clearly have may other potential uses.
These projects are all based on sub-samples of {\it ROSAT}-selected clusters
to minimize the number of observations required. The reluctance
of time allocation committees to devote time to complete samples
in preference to the ``exotica'' (e.g., most distant, strongly lensing,
etc., which predominate in successful proposals) is a hindrance to
this ``targeted'' survey approach.

\begin{figure*}[t]
%\hspace{-1.5cm}
\includegraphics[width=0.75\columnwidth,angle=270,clip]{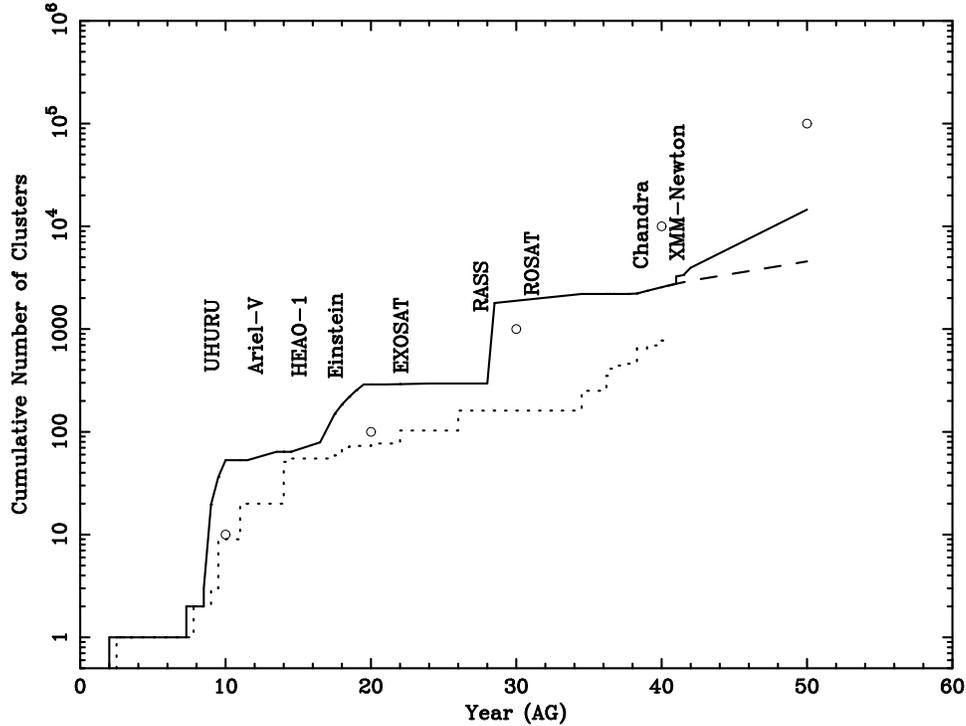}
\vskip 0pt \caption{
The total number of clusters with X-ray detections
known with time. The solid line marks the number detected
or likely to be detected. The dotted line marks the detections
published in the literature. The continuation above 40~AG is
for an optimistic (solid) and pessimistic (dashed) assumption
for the efficiency of {\it XMM-Newton} serendipitous surveys. The
circles mark the number required for ``Edge's Law'' to hold.
\label{Lx-redshift}}
\end{figure*}

\section{Can ``Edge's Law'' Hold?}

The simple answer to this is question ``No.'' Why should it?
The observable Universe is finite so the number of clusters
cannot grow exponentially indefinitely.
Figure~1.2 shows the final version of the plot I showed
during my talk with the cumulative number of clusters
known with time. This illustrates the recent slowing of
the number of clusters discovered and the increased lag between
the detection and final publication of clusters.

As Simon White pointed out at the end of my talk, the
case for ever-increasing sensitivity for the sake of it
is a poor foundation for any field. The source counts
for clusters crudely imply that every order
of magnitude increase in number translates to an
order of magnitude better sensitivity. So for the 
present, year 40~AG, the  ``Edge's Law'' requirement
would be equivalent to an all-sky survey to a flux
of $10^{-13}$ erg~s$^{-1}$~cm$^{-2}$ (0.5--2.0~keV).
When I retire this will be $10^{-15}$ erg~s$^{-1}$~cm$^{-2}$ (0.5--2.0~keV).
This flux limit is reachable with current missions, but less
than 1\% of the sky could be covered. 

There are, however, strong arguments for larger, deeper
contiguous X-ray surveys than are available now or in the
near future. The cosmological constraints that can be derived
from the large-scale clustering of clusters, their mass function, 
and chemical evolution are complementary to those available
elsewhere, most notably {\it WMAP} (Spergel et al. 2003). With the
existence of large-area, multicolor optical and near-infrared
surveys (e.g., SDSS, CHFTLS, RCS2, UKIDSS, Vista), the bottleneck
of identification is eased and photometric redshifts will be
sufficient for most purposes. 

On a more practical level, the next generation of X-ray satellites,
{\it XEUS} and {\it Constellation-X}, are optimized for the study of faint
($10^{-14}$ erg~s$^{-1}$~cm$^{-2}$) sources. To get the best
from the massive investment made in these missions, it would
be sensible to have surveyed more than 1\% of the sky to this
depth. 

Several proposals have been made to do this and are listed in Table~1.3.  To 
date none of these missions is fully approved.  The case for {\it DUET} was 
based on a survey of the SDSS area (Jahoda et al. 2003)
and would have detected 20,000 clusters. It is likely that a proposal
of this type will succeed (probably {\it ROSITA}) so some form of all/part-sky
survey will have been performed by 50--55~AG.
It is very unlikely that any mission (proposed or yet to be proposed) is likely
to reach the limits required to keep above ``Edge's Law,'' but
the constant progress made in X-ray astronomy will see the
number of clusters increase to well above 30,000 by 60~AG. 
This should be sufficient for cosmological work and more
detailed studies with the next generation of X-ray satellites.

\begin{table}
\caption{X-ray Missions}
\begin{tabular}{lll}
\hline
Mission    & Country & Status \\
 & & \\ \hline
{\it ABRIXAS-II} & German & renamed {\it ROSITA} \\
{\it WFXT}      & Italian           & rejected \\
{\it PANORAM-X} & ESA Flexi Mission & rejected \\
{\it ROSITA}    & ESA ISS Mission & accepted phase A, could fly 2007 \\
{\it DUET}      & NASA Pathfinder & rejected \\ \hline
\end{tabular}
\end{table}

\section{Conclusions}

The success of X-ray surveys in the selection and understanding
of clusters of galaxies over the past four decades has been remarkable.
The statistical properties of X-ray selected samples of clusters
have been used to determine cosmological parameters, and the detail
found in individual clusters can be used to understand the evolution of
that cluster. Current and planned surveys will build on these 
previous studies and will undoubtedly reveal further 
complexity in the intracluster medium, thereby refining our
understanding of the astrophysics of these systems.

\section{A Coda}

As some of you will know, I was only able to attend the conference
for one day due to the death of my father, David Edge. Fewer of you will know
that one of my father's many legacies is one of the cornerstones
of modern astronomy, the 3rd Cambridge Radio Catalogue (3C), which
was his Ph.D. thesis. He always kept a keen interest in astronomy
but moved on to become a leading figure in the sociology of science.
I would like to thank Keith Taylor for his help on the evening
before my departure and Richard and Barbara Ellis
for providing the venue. Thanks
too to all my friends and colleagues who have contacted me since
then.

\vspace{0.3cm}
{\bf Acknowledgements}.
I owe Dave Gorman thanks for providing me with a presentation format
that kept the audience awake. I am grateful to all those with whom I have 
worked on X-ray surveys, but particular thanks go to Harald Ebeling, whose
contribution to the field is unrivaled.

\end{document}